\documentclass[a4paper]{jpconf}
\usepackage{graphicx}
\usepackage{pifont}
\usepackage[square,sort,comma,numbers]{natbib}
\usepackage{hyperref}
\usepackage{verbatim}
\hypersetup{
colorlinks = true,
urlcolor   = blue,
linkcolor  = blue,
citecolor  = blue
}
\usepackage{comment}
\begin{document}
\title{Magnetic Properties of Multifunctional $^7$LiFePO$_4$ under Hydrostatic Pressure}

\author{Ugne Miniotaite$^{1,*}$, Ola Kenji Forslund$^2$, Elisabetta Nocerino$^1$, Frank Elson$^1$, Rasmus Palm$^1$, Nami Matsubara$^1$, Yuqing Ge$^2$, Rustem Khasanov$^3$, Genki Kobayashi$^{4,5}$, Yasmine Sassa$^2$, Jonas Weissenrieder$^1$, Vladimir Pomjakushin$^6$, Daniel Andreica$^7$, Jun Sugiyama$^8$, and Martin M{\aa}nsson$^{1,+}$}
\address{$^1$Sustainable Materials Research \& Technologies (SMaRT), Department of Applied Physics, KTH Royal Institute of Technology, SE-106 91 Stockholm, Sweden}
\address{$^2$Department of Physics, Chalmers University of Technology, Göteborg, SE-412 96, Sweden}
\address{$^3$Laboratory for Muon Spin Spectroscopy, Paul Scherrer Institute, CH-5232 Villigen PSI, Switzerland}
\address{$^4$Solid State Chemistry Laboratory, Cluster for Pioneering Research (CPR), RIKEN, 2-1 Hirosawa, Wako, Saitama 351-0198, Japan}
\address{$^5$Department of Materials Molecular Science, Institute for Molecular Science, National Institutes of Natural Sciences, 38 Nishigonaka, Myodaiji, Okazaki 444-8585, Japan}
\address{$^6$Laboratory for Neutron Scattering \& Imaging, Paul Scherrer Institute, CH-5232, Villigen PSI, Switzerland}
\address{$^7$Faculty of Physics, Babes-Bolyai University, 400084 Cluj-Napoca, Romania}
\address{$^8$Neutron Science and Technology Center, Comprehensive Research Organization for Science and Society (CROSS), Tokai, Ibaraki 319-1106, Japan}

\ead{$^{*}$ugnem@kth.se, $^{+}$condmat@kth.se}
\begin{abstract}
LiFePO$_4$ (LFPO) is an archetypical and well-known cathode material for rechargeable Li-ion batteries. However, its quasi-one-dimensional (Q1D) structure along with the Fe ions, LFPO also displays interesting low-temperature magnetic properties. Our team has previously utilized the muon spin rotation ($\mu^+$SR) technique to investigate both magnetic spin order as well as Li-ion diffusion in LFPO. In this initial study we extend our investigation and make use of high-pressure $\mu^+$SR to investigate effects on the low-$T$ magnetic order. Contrary to theoretical predictions we find that the magnetic ordering temperature as well as the ordered magnetic moment increase at high pressure (compressive strain).  
\end{abstract}

\section{Introduction}
Multifunctional materials are defined as materials that have several interesting and applicable properties. For example, a material can both be a superconductor as well as an insulator, it is just a question of which temperature range the material is operated in. In the same way, the LiFePO$_4$ (LFPO) compound is both a cathode material for rechargeable batteries \cite{Padhi_1997,Yamada_2001,Kobayashi_2009,Sugiyama_2011,Zhang_2011,Sugiyama_2012,Sugiyama_2012_2} in a room temperature regime, while displaying interesting magnetic properties at low temperatures \cite{Santoro_1967,Sugiyama_2011,Sugiyama_2012_2,Ofer_2012,ToftPetersen_2015,Werner_2021}. Originally, the great interest for LFPO is founded in its electrochemical properties, where it is of high interest as a safe and highly stable cathode material in low current conditions (c.f. other Li-ion battery cathode materials) \cite{Mohamed_2020}. Further, LFPO is a 'Co-free' cathode material where the replacement of Co with Fe is very favorable due to its much higher natural abundance (Fe:Co = 2500:1). Consequently, both lower price as well as less friction on the geopolitical level makes LFPO attractive for industrial scale energy applications.

\begin{figure}[ht]
  \begin{center}
    \includegraphics[scale=1.34]{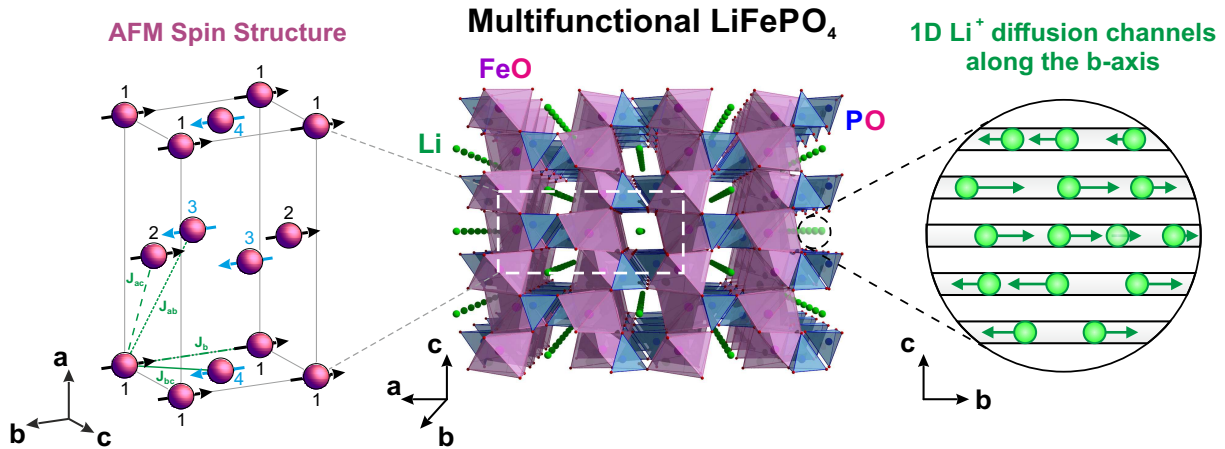}
  \end{center}
  \caption{Atomic model of LFPO with zoomed in view on the 1D Li-ion diffusion channels along the b-axis (right) as well as the antiferromagnetic (AFM) spin structure of four different sublattices along with main magnetic interactions $J$ (left, adopted from Refs.~\cite{Werner_2021,ToftPetersen_2015} that contain further details).}
  \label{LFPO}
\end{figure}

LFPO adopts an olivine type crystal structure with the orthorhombic $Pnma$ space group (see Fig.~\ref{LFPO}). The Li ions are situated in one-dimensional (1D) channels running along the b-axis \cite{Nishimura_2008}, which give rise to interesting effects on the lithium intercalation mechanism and dynamics \cite{Malik_2013}. Such anisotropic intercalation/de-intercalation process during charge/discharge cycles is generally believed to occur topotactically, and thereby create a phase separated cathode material with LFPO and FPO phases, respectively. However, such two-phase separation mechanism has also been called into question by both theoretical \cite{Bai_2011} as well as experimental studies \cite{Gu_2011}. At low temperature, LFPO enters into an antiferromagnetic (AFM) phase below $T_{\rm N}\approx50$~K \cite{Santoro_1967}. It has been shown that a rather complex canted spin order is present \cite{ToftPetersen_2015} (see also Fig.~\ref{LFPO}) resulting from competing in-plane and out of plane magnetic couplings ($J$).

A lot of effort has already been done in order to tune the ion-dynamics (i.e. battery performance) in LFPO. Here, nanostructuring as well as surface coatings (with carbon as well as with other elements and compounds) \cite{Dominko_2005} are among the most common approaches. Such systems have also been studied systematically using both neutron scattering \cite{Benedek_2019} and $\mu^+$SR \cite{Benedek_2020}. When it comes to the tuning of low-temperature magnetic properties, much less detailed studies are to be found in the published literature. Indeed, chemical substitution in LFPO is possible, yielding e.g. NaFePO$_4$, which display a slightly larger unit cell \cite{Avdeev_2013} and also an expected reduction in the magnetic ordering temperature ($T_{\rm N}$) by a few Kelvin (Fe-Fe distance is larger). Of course, nanostructuring and surface coatings could also induce interesting effects on the magnetic properties, however, other more direct and 'cleaner' approaches could be employed first. In fact, a magnetic field dependence of the AFM magnons in LFPO has recently been presented \cite{Werner_2021}. Another very common method is to apply hydrostatic pressure, thereby compressing the crystallographic lattice, and investigate how this affects the spin correlations in a material. Combining the penetrating properties of muons, the ultra-high sensitivity of $\mu^+$SR to detect changes in spin order/dynamics, along with hydrostatic pressure is a very powerful approach \cite{Thede_2014,Forslund_2019,Sugiyama_2020,Sugiyama_2021}. In this study we have used a similar protocol to conduct the first investigation of the local magnetic properties of LFPO under hydrostatic pressure.

\section{Experimental Setup}
The $^7$LiFePO$_4$ powder sample was prepared using a solid-state reaction technique with $^7$Li enriched and reagent grade $^7$Li$_2$CO$_3$, Fe(II)C$_2$O$_4\cdot$2H$_2$O, and (NH$_4$)2HPO$_4$ as the starting materials. More details on the sample synthesis can be found in Ref.~\cite{Sugiyama_2011}.

The neutron powder diffraction (NPD) experiments were performed using the HRPT instrument at the Swiss Spallation Neutron Source (SINQ) of Paul Scherrer Institute (PSI), Switzerland. Approximately 2 grams of sample inside a standard {\O} = 6 mm vanadium sample container that was subsequently inserted into a cryofurnace ($1.5 K\leq{}T\leq550$~K). High resolution data was acquired using the neutron wavelength $\lambda=1.49$~{\AA}. The NPD patterns were refined using the Fullprof software suite \cite{Fullprof_ref}.

The $\mu^+$SR experiments were performed at $\mu$E1 beamline within the S$\mu$S muon source of PSI using the GPD instrument \cite{Khasanov_2016,Khasanov_2022}. Hydrostatic pressures up to 20 kbar were applied utilizing a double-wall piston cylinder cell (MP35/CuBe alloy). Daphne oil was used as the pressure medium in order to achieve the hydrostatic pressure. Measurements were performed inside a $^{4}$He flow cryostat, reaching temperatures down to 2~K. More details about the $\mu^+$SR technique \cite{muSRbook_1,muSRbook_2} and experimental setup can be found in Refs.~\cite{Khasanov_2016,Khasanov_2022,Andreica}. Data was analyzed using the open analysis software suite \texttt{musrfit} \cite{musrfit}. 

\begin{figure}[ht]
  \begin{center}
    \includegraphics[scale=0.7]{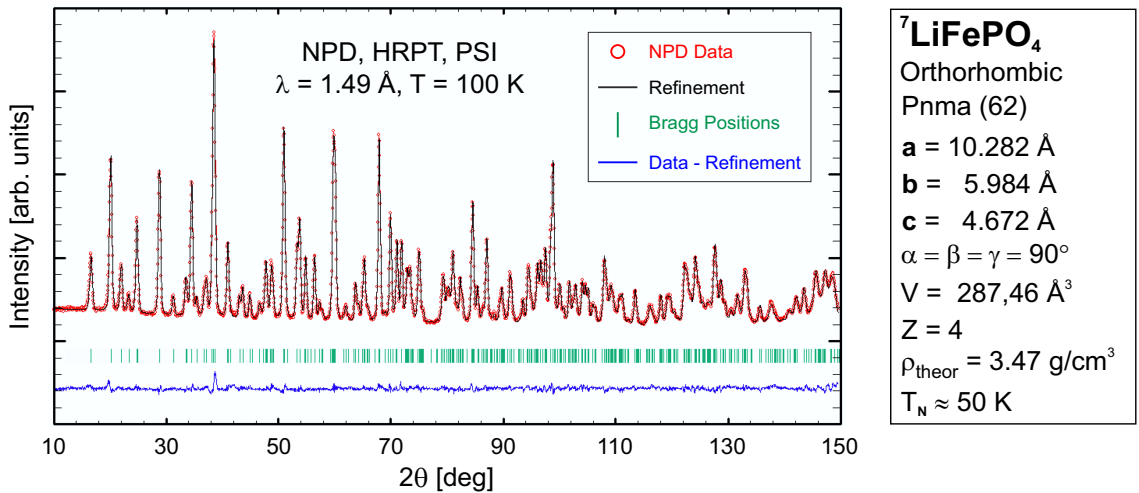}
  \end{center}
  \caption{Neutron powder diffraction (NPD) pattern from $^7$LiFePO$_4$ acquired using a neutron wavelength $\lambda=1.49$~{\AA}. The right panel show crystallographic data obtained from the Rietveld refinement of our NPD data at $T=100$~K. $V$ describing the volume of the unit cell, $Z$ the number of molecules per unit cell, $\rho_{\rm theor}$ the theoretical density of the compound, and $T_{\rm N}$ the Néel temperature.}
  \label{NPD}
\end{figure}

\section{Results \& Discussion}
Initial neutron powder diffraction (NPD) investigation show that the sample is of high quality. As shown in Fig.~\ref{NPD}, the diffraction pattern acquired at $T=100$~K is very well refined by the expected $Pnma$ space group and there are no signs of any impurity peaks. The right panel of Fig.~\ref{NPD} further display the basic crystallographic data obtained from the Rietveld refinement.

\begin{figure}[ht]
  \begin{center}
    \includegraphics[scale=1.2]{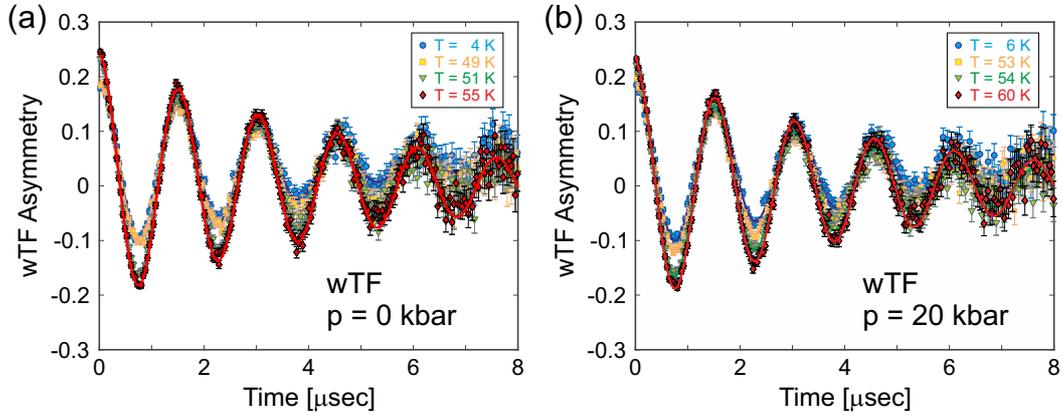}
  \end{center}
  \caption{wTF $\mu^+$SR spectra of LFPO at selected temperatures for (a) ambient pressure and (b) $p=20$~kbar, respectively. Solid lines are fits to the data using Eq.~\ref{wtf}.}
  \label{wTFspec}
\end{figure}

Weak transverse-field (wTF~=~50~G) $\mu^+$SR time spectra were collected as a function of temperature for selected pressures ($p=0, 10, 20$~kbar). Typical wTF time spectra at selected temperatures are shown in Fig.~\ref{wTFspec} for ambient and $p=20$~kbar, respectively. A distinct oscillation is observed as well as an offset at lower temperatures. Therefore, wTF data was fitted using a combination of an exponentially relaxing cosine function and an exponentially relaxing signal:

\begin{eqnarray}
 A_0 \, P_{\rm TF}(t) = A_{\rm TF}\cos(2\pi f_{\rm TF}t + \phi)e^{-\lambda_{\rm TF} t}+A_{\rm S}e^{-\lambda_{\rm S} t},
\label{wtf}
\end{eqnarray}

where $A_0$ is the initial asymmetry, which depends on the instrumental configuration, and $P_{\rm TF}(t)$ is the muon spin depolarisation function under TF. $A_{\rm TF}$, $f_{\rm TF}$, $\phi$ and $\lambda_{\rm TF}$ are the asymmetry, frequency, phase and relaxation rate originating from the applied TF while $A_{\rm S}$ and $\lambda_{\rm S}$ are the asymmetry and relaxation rate originating from the internal magnetic field. 

\begin{figure}[ht]
  \begin{center}
    \includegraphics[scale=1.3]{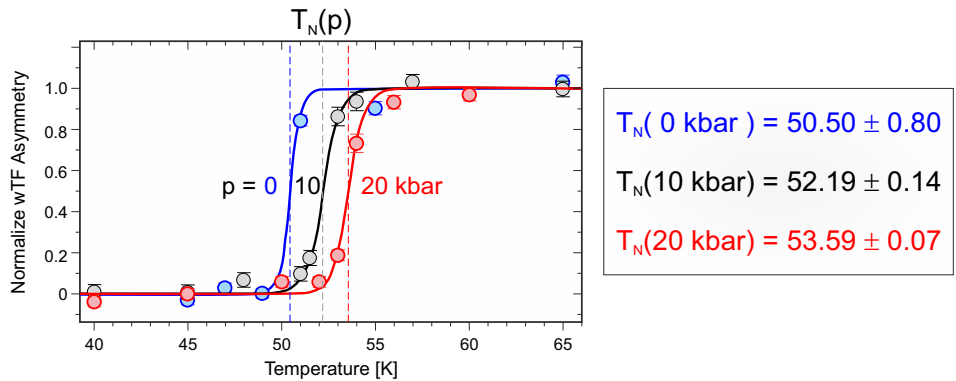}
  \end{center}
  \caption{wTF asymmetry at $t=0$ as a function of temperature and pressure. Solid line represent a fit to a sigmoid type function. Note that the asymmetry has been background subtracted and normalized to unity following the fit to the sigmoid function. The resulting values for $T_{\rm N}(p)$ are shown in the right panel.}
  \label{wTFAsym}
\end{figure}

Fig.~\ref{wTFAsym} shows the background subtracted and normalized $A_{\rm TF}$ as a function of temperature for the three different pressures. For all pressures, $A_{\rm TF}$ exhibits a temperature independent behaviour at low temperatures. A step-like increase is observed for each pressures as the temperature increases. Since $A_{\rm TF}$ corresponds roughly to the paramagentic fraction, the step like increase is attributed to the magnetic transition. This critical temperature is obtained from a sigmoid fit and is defined as the point in which the paramagnetic fraction $\simeq$ magnetic fraction function for each pressure [$T_{\rm N}(p)$]. A clear pressure dependence is observed in $T_{\rm N}$, which increases with applied hydrostatic pressure. Previous inelastic neutron scattering measurements determined most significant bonds with values: $J_1=-1.08$~meV, $J_2=-0.4$~meV and $J_\perp=0.021$~meV \cite{Jiying2006} where $J_1$ and $J_2$ are in plane interactions along [001] and [010] while $J_\perp$ is the perpendicular interaction along [100]. Moreover, high pressure XRD at room temperature reported a higher compressibility along the [100] direction (a-axis is the longest) \cite{Dong2017}. Therefore, the increase in $T_{\rm N}(p)$ with pressure may be attributed to stronger $J_\perp$ interactions due to shorter bond distance, at least as a first approximation. It is noted that $J_\perp$ is a ferromagnetic interaction, unlike $J_1$ and $J_2$. Therefore, a high pressure AF-FM transition may be present in LFPO at low temperatures. 

\begin{figure}[ht]
  \begin{center}
    \includegraphics[scale=1.05]{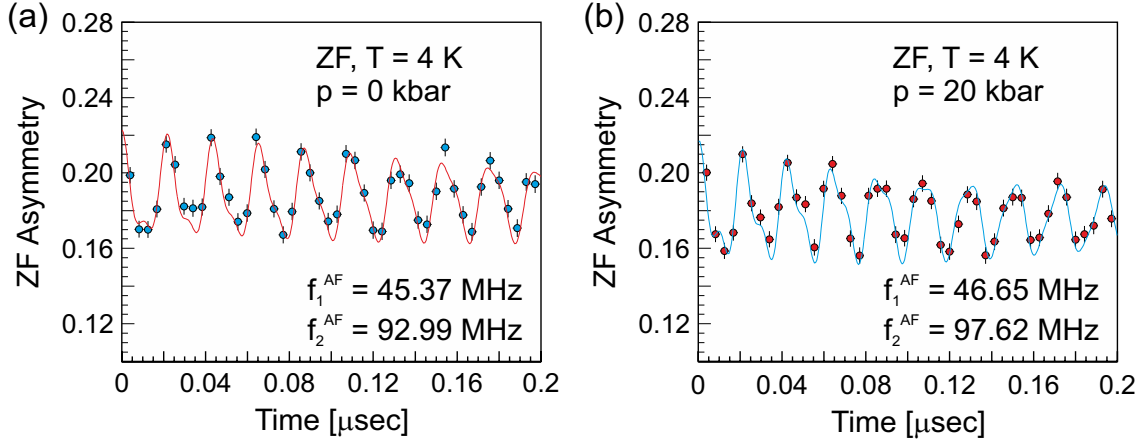}
  \end{center}
  \caption{Zero field (ZF) $\mu^+$SR spectra acquired at $T=4$~K and (a) ambient pressure, as well as (b) $p=20$~kbar. Solid line is the fit to Eq.~\ref{zf}, resulting in the two frequencies $f^{\rm AF}_1$ and $f^{\rm AF}_2$.}
  \label{ZFspec}
\end{figure}

Zero field (ZF) time spectra were collected at base temperature $T=4$~K for ambient and highest (20 kbar) pressures. The spectra exhibit two distinct oscillations [see Fig.~\ref{ZFspec}(a)], consistent with our previous study \cite{Sugiyama_2011}. Therefore, they were fitted using a combination of exponentially relaxing cosine functions, an exponential and a exponentially relaxing static gaussian Kubo-Toyabe ($G^{\rm SGKT}$):

\begin{eqnarray}
 A_0 \, P_{\rm ZF}(t) &=& \sum_i^2 A^{\rm AF}_{i}\cos(f^{\rm AF}_{i}t + \phi^{\rm AF}_{i})e^{-\lambda^{\rm AF}_{i} t}+ A_{\rm tail}e^{-\lambda_{\rm tail} t}+ A_{\rm PC}G^{\rm SGKT}e^{-\lambda_{\rm PC} t}
\label{zf}
\end{eqnarray}

where $A_0$ is the initial asymmetry, which depends on the instrumental configuration, and $P_{\rm ZF}(t)$ is the muon spin depolarisation function under ZF. $A^{i}_{\rm i}$, $f^{\rm AF}_{\rm i}$, $\phi^{\rm AF}_i$ and $\lambda^{\rm AF}_{\rm i}$ are the asymmetry, frequency and phase resulting from internal field contributions that are non-parallel to the initial muon spin polarisation. $A_{\rm tail}$ and $\lambda_{\rm tail}$ are the asymmetry and relaxation rate of the tail component, that originates from internal field components that are parallel to the initial muon spin polarisation. $A_{\rm PC}$ and $\lambda_{\rm PC}$ are the asymmetry and relaxation rates, respectively, associated with the $G^{\rm SGKT}$ signal, attributed to contributions from the pressure cell \cite{Forslund_2019}.  

\begin{figure}[h]
  \begin{center}
    \includegraphics[scale=1.05]{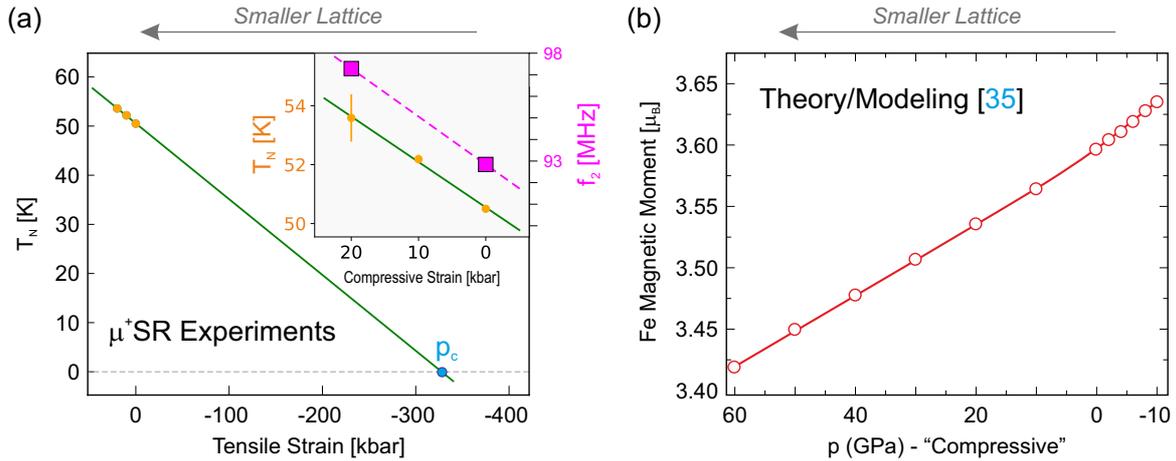}
  \end{center}
  \caption{(a) Pressure dependence of $T_{\rm N}$ as well as the ZF $\mu^+$SR frequency $f_2$ (inset). A critical pressure can be estimated as $p_{\rm c}\approx-325$~kbar. (b) Theoretical prediction for the Fe ordered magnetic moment in LiFePO$_4$ for the $Cmcm$ structure (adopted from Ref.~\cite{Lin_2011}). Note that the x-axes have been selected to intuitively show a decrease in the lattice volume rather than as an increase in applied pressure.}
  \label{pdep}
\end{figure}

The precession frequencies at ambient pressure, $f^{\rm AF}_1=45.37$~MHz and $f^{\rm AF}_2=92.99$~MHz are fully consistent with our previous studies \cite{Sugiyama_2011}. When a hydrostatic pressure $p=20$~kbar is applied (i.e. compression of the lattice), the value of both $f^{\rm AF}_1$ and $f^{\rm AF}_2$ increases (see Fig.~\ref{ZFspec}), which is coherent with the TF results showing an increase in $T_{\rm N}$. Such results are summarized in Fig.~\ref{pdep}(a). An increase in the values of precession frequency, implies an increase in the ordered moment and supports a scenario in which $J_\perp$ increases in strength with pressure application. This is also very reasonable when considering the NaFePO$_4$ sister compound mentioned above, which displays a larger Fe-Fe distance, and as a result, slightly lower $T_{\rm N}$ \cite{Avdeev_2013}. Several high pressure studies of the atomic structure in LFPO have been published and there is some controversy in the results. An early high pressure XRD investigation reported a structural transition to a space group $Cmcm$ at a critical pressure below 65 kbar \cite{Garc2001}. However, other studies \cite{Valdez_2018} indicate a much higher critical pressure. Further, theoretical calculations based on the $Cmcm$ structure predict a decrease in the ordered Fe magnetic moment \cite{Lin_2011} with applied pressure, as shown in Fig.~\ref{pdep}(b). Our experimental result on the contrary show stronger magnetic correlations with applied pressure. Therefore, we may at least conclude that the supposed structural transition is not present up to 20~kbar. Our TF and ZF result also indicate that the spin structure at ambient pressure is most likely preserved up to at least $p=20$~kbar. We can of course not fully exclude a combination of atomic lattice compression together with a subtle change in the spin canting occurs. However, to discern such in-depth information, it would be necessary to conduct additional and detailed high-pressure neutron diffraction studies.

Finally, even if we only have acquired data at three different pressures, we could indeed fit both $T_{\rm N}(p)$ and $f^{\rm AF}(p)$ to a straight line. For $T_{\rm N}(p)$ we can then extrapolate and acquire a critical pressure $p_c$ for which $T_{\rm N}=0$~K, i.e. magnetic order in LFPO is fully suppressed. Such procedure yield $p_c\approx-325$~kbar, i.e. a very large tensile strain (larger lattice). This is clearly a very unreasonable phase space for practically conducting actual experiments, however, it gives an indication and guidance to how LFPO could be chemically tuned in order to suppress magnetic spin order.

\section{Conclusion}
We have performed the first $\mu^+$SR investigations of the magnetic properties in LiFePO$_4$ as a function of hydrostatic pressure and temperature. Measurements in transverse field (TF) and zero field (ZF) confirms the formation of long range antiferromagnetic order within the entire pressure range ($p=0-20$~kbar). A clear increase in both magnetic ordering temperature ($T_{\rm N}$) and ordered Fe magnetic moment is revealed, which we interpret as related to an increase of the out plane interaction strength. Moreover, our results also confirm the absence of any $Pnma$ to $Cmcm$ structural transition in LiFePO$_4$, at least up to $p=20$~kbar. Finally, we can estimate a critical pressure $p_{\rm c}\approx-325$~kbar (i.e. tensile strain) for which magnetic spin order in LiFePO$_4$ is fully suppressed.

\section{Acknowledgments}
The authors wish to thank the staff of S$\mu$S and SINQ (PSI) for their great support during the experiments. This research is funded by the School of Science and Engineering (SCI) of the KTH Royal Institute of Technology, the Ragnar Holm Foundation, the Swedish Foundation for Strategic Research (SSF) within the Swedish national graduate school in neutron scattering (SwedNess), as well as the Swedish Research Council VR (Dnr. 2021-06157 and Dnr. 2017-05078), and the Carl Tryggers Foundation for Scientific Research (CTS-18:272). J.S. was supported by the Japan Society for the Promotion Science (JSPS) KAKENHI Grant No.No. JP18H01863 and  JP20K21149. D.A. acknowledges financial support from the Romanian UEFISCDI project PN-III-P4-ID-PCCF-2016-0112, Contract Nr. 6/2018. Y.S. and O.K.F. are also funded by the Chalmers Area of Advance - Materials Science.

\bibliographystyle{iopart-num}
\bibliography{refs}

\end{document}